# Critical behavior of the specific heat in Ti-Si amorphous alloys at the metal-insulator transition

A. Rogachev[1], H. Ikuta[2], and U. Mizutani[3]

[1]Department of Physics and Astronomy, University of Utah, Salt Lake City, 84112 USA
[2]Department of Materials Physics, Nagoya University, Nagoya 464-8603, Japan
[3] Nagoya Industrial Science Research Institute, Nagoya, Japan

**Abstract**

In this paper, we report the measurements of specific heat of an amorphous $Ti_{9.5}Si_{90.5}$ alloy located very close to the critical point of the metal-insulator transition. In the presence of a magnetic field, the specific heat is dominated by the Schottky anomaly caused by magnetic moments associated with the dangling bonds in the matrix of amorphous Si. Subtraction of this contribution exposes the behavior of the electronic specific heat coefficient $\gamma$. The coefficient is temperature-independent above 2 K and is, in order of magnitude, close to the value expected in the absence of electron-electron interactions. In the temperature range 0.4-1.5 K, the coefficient $\gamma$ shows an anomalous downturn, which can be approximated by the dependence $\gamma(T) = \gamma_0 \ln(T/T_0)$, with $T_0 \approx 0.2$ K . In a companion paper, we found that the Hall coefficient in Ti-Si alloys is affected by the electron-electron interaction up to much higher temperature of 150 K and also varies critically across the metal-insulator transition. We compare our results with theoretical predictions for three models, which can potentially explain the anomalous behavior of the specific heat: generalized non-linear $\sigma$ model, Coulomb glass, and many-body localization.

*Introduction.*

Systems with strong disorder and interactions present one of the grand challenges in condensed matter physics [1]. The study of these systems has a long history marked by many still unresolved problems. In particular, unlike the case of several clean systems [2,3,4], the connection between the long-range behavior that determines a system response near a metal-insulator transition (MIT) and the corresponding microscopic physics has not been established. For example, it is not clear why the correlation length exponent, $v$, and dynamical exponent, $z$, are different in crystalline silicon doped with phosphorous (Si:P, $v = 1$, $z = 3$) [5] and with boron (Si:B, $v = 1.6$ , $z = 2$) [6]. Nor is it known why the experimental exponents for Si:B coincide with values predicted for a non-interacting disordered 3d system, in clear conflict with the tunneling experiments, which indicate the emergence of the Coulomb gap at the critical point of MIT [7]. For the insulating state, there is a long-standing question of what the nature of the current carriers is. Experimentally, the large body of the data on conductivity can be explained by Efros-Shklovskii variable-range hopping (VRH), based on the picture that current is carried by single-particle hops in the presence of the Coulomb gap. Mott and Pollak [8], however, argued that in the presence of interactions, a jump of an individual electron has to lead to the displacement of neighboring elections, so the true carriers are many-electron entities called dressed polarons. Extending this reasoning to thermodynamic quantities, one can see that the relaxation of the whole system requires a complicated re-arrangement of a gigantic number of elections, resulting in a dynamically frozen state known as a Coulomb glass. Contrary to these expectations, several measurements of specific heat in doped semiconductors and amorphous alloys revealed a simple smooth variation of the electron coefficient γ across MIT [9,10,11,12].

A modern approach to the problem of interacting particles in a strongly disordered potential takes a broader perspective by also considering systems in their excited states and completely isolated from their environment. There is strong evidence in these systems for the existence of a new, dynamical state of matter caused by many-body localization (MBL) [13,14], which does not thermalize and retains the memory of its initial state. It was first theoretically discovered in zero-dimensional systems [15] and one-dimensional systems with local interactions [16,17]. Recent theoretical works [18,19] and experiments on cold atoms [20,21] suggest that MBL might exist in 2d and 3d systems, even in the case of a long-range $1/r$ interaction [22]. The MBL state is also of great interest for standard condensed matter systems. Recently, it has apparently been observed in highly disordered InO films [23]. On the theoretical side, significant effort has been made to analyze MBL in systems weakly connected to the environment [24,25,26,27]. One conclusion is that, although these systems eventually equilibrate, this process is logarithmically slow and implies exponentially large relaxation time and a possibility to see the MBL state in experiments.

In the present work, we report an observation of the critical behavior of the electronic specific heat coefficient $\gamma$ in amorphous $Ti_{9.5}Si_{90.5}$ alloy located right at the critical concentration of the metal-insulator transition. We have observed a sharp logarithmic decrease of γ below about 1.5 K that extrapolates to zero at 0.2 K. We argue that this behavior might reflect the emergence of the Coulomb glass or MBL state in the system.

*Experiment.*

The $Ti_{9.5}Si_{90.5}$ sample was fabricated in a DC sputtering apparatus custom-designed for the production of bulk samples. It was one of a series of samples used previously for specific heat [12], transport [12,28] and magnetic [29] measurements. The sputtering targets for this series were made in-house by arc-melting of high-grade Ti (99.99 %) and Si (99.999 %). The uniformity of the targets was checked by measuring the near-surface composition of the targets by energy dispersive x-ray analysis (EDXA) before and after sputtering. The sample holder was made out of two parts. The first part was a round (diameter 1 inch) copper gasket that was screwed vacuum-tight on a stainless-steel hollow rod that could be grounded or DC biased with high voltage. The second part was a thin copper plate with a matching diameter and thickness 0.3-0.5 mm; it was soldered to the platform with indium. A fresh copper plate was used in every deposition; it was cleaned by back-sputtering and served as a substrate. The sample holder was water-cooled during deposition. The sputtering was done in *tour de force* fashion. A short, 5 cm, distance between the target and sample provided high deposition rate of 2-4 nm/sec. The process was typically run for more than 24 hours. As a result, we could deposit material with thickness 100-200 μm. After the deposition, the sample gasket was removed from the machine and heated up to let the indium melt so the copper plate could be detached from the gasket. Then, the plate was gently bent in different directions; this allowed brittle Ti-Si samples to be peeled off the copper plate. Typically, the obtained samples were in a powder or needle-like form. For $Ti_{9.5}Si_{90.5}$ alloy, however, we were fortunate to get a collection of large pieces (lateral size of about 5 mm) suitable for specific heat measurements down to 0.3 K. We used EDXA to check the composition of the top and bottom surface of the samples; we found that there is no copper contamination and estimated that the uniformity of the sample is within 1 at. % of Ti.

Transport properties of the sample were measured with a Quantum Design PPMS in the temperature range 1.8-300 K and magnetic field up to 9 T. Specific heat measurements were done using the relaxation method in Oxford Instruments MLHC9H, calibrated as prescribed by the vendor in the temperature range 0.4-10 K and magnetic fields up to 6 T. Several pieces of Ti-Si sample were mounted with Apiezon grease on a suspended sapphire chip equipped with a heater and temperature sensor. Standard procedures to measure and subtract the bare chip heat capacity were used. Unfortunately, we do not have the resistance data in the temperature range 0.4-1.8 K. They were not taken at the time when the specific heat was measured and when, after seven years, we came back to this project, we found that samples had changed and become significantly more resistive.

*Results and discussion.*

The conductivity of the $Ti_{9.5}Si_{90.5}$ amorphous alloy is shown in Fig. 1(a) as a function of the square root of temperature. As the linear fit presented in the figure shows, in the temperature range 1.8-10 K, the conductivity can be nicely fitted with a simple expression $\sigma(T) = \sigma(0) + \alpha T^{1/2}$, with $\sigma(0) = 1.5\ \Omega^{-1}cm^{-1}$ and $\alpha = 8.5\ \Omega^{-1}cm^{-1}K^{-1/2}$. The temperature-dependent term has the form of the quantum correction due to electron-electron interaction (EEI) for a 3-dimensional system. The same behavior was observed on the metallic side of the MIT in the measurements on doped semiconductors [5,6] and amorphous alloys [30] extended to much lower temperatures. While we cannot exclude a possibility that at lower temperatures the conductivity of the $Ti_{9.5}Si_{90.5}$ alloy switches to the insulating behavior, it is clear that the sample is located almost at the critical point of MIT.

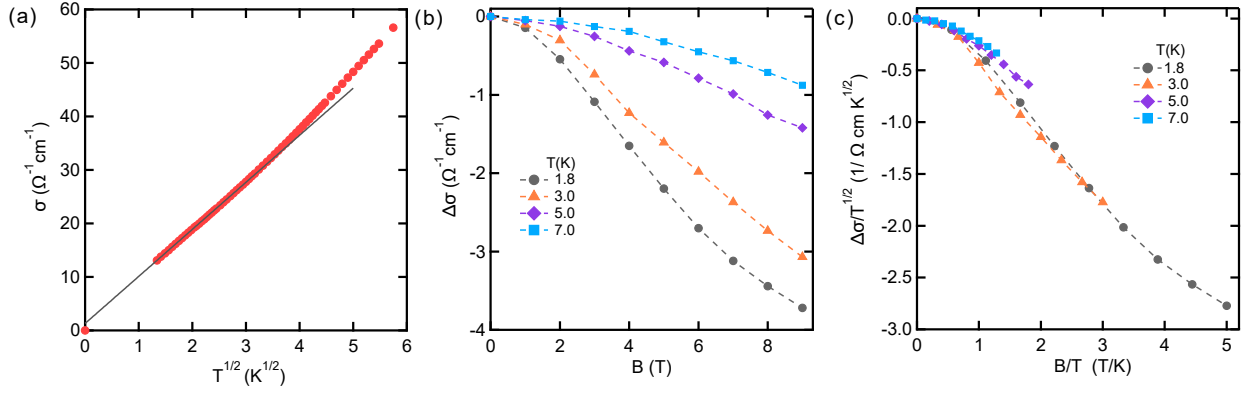

*Fig. 1. (a) Conductivity of the $Ti_{9.5}Si_{90.5}$ amorphous alloy as function of the square root of temperature. (b) Magnetoconductivity, $\Delta\sigma(T,B) = \sigma(T,B) - \sigma(T,0)$, as a function of magnetic field at several indicated temperatures. (c) Scaled magnetoconductivity versus scaled magnetic field.*

We have also carried out transport measurements of the alloy in magnetic fields. The results are presented in Fig. 1(b) in the form of magnetoconductivity defined as $\Delta\sigma(T,B) = \sigma(T,B) - \sigma(T,B=0)$. The magnetic field produces a fairly large effect on conductivity, reducing it by about 30 % at $T = 1.8$ K and $B = 9$ T. Surprisingly, the observed behavior of $\Delta\sigma(T,B)$ matches fairly closely the prediction made by the perturbation theory of the EEI correction [31]. Functionally, it has the form $\Delta\sigma(T,B) = a_1 T^{1/2} g_3(h)$, where $h = a_2 B/T$, $a_1$ and $a_2$ are some constants that depend on the parameters of the alloy, and the function $g_3(h)$ has the limited behavior: $g_3(h) = \sqrt{h} - 1.3$ at $h \gg 1$ and $g_3(h) = 0.053 h^2$ at $h \ll 1$. This is indeed the behavior we see at high and low fields. From the formula, it also follows that magnetoconductivity data collapse on a single curve when plotted as $\Delta\sigma(B,T)/T^{1/2}$ versus $B/T$. The scaled data shown in Fig. 1(c) come close to this variation. Overall, it appears that above 1.8 K, the alloy displays metallic behavior dominated by the EEI quantum correction.

The specific heat of the alloy measured at the indicated magnetic field is presented in Fig. 2, in the form $C/T$ versus $T^2$. For a nonmagnetic metal, the low-temperature specific heat is expected to follow the equation $C = \gamma T + \beta T^3$. This is the behavior we see for the alloy in zero magnetic field at the temperatures above about 2 K (Fig. 2a). The intercept gives the electronic coefficient $\gamma$ and the slope defines the phonon contribution $\beta$. They closely agree with the previous measurements [12].

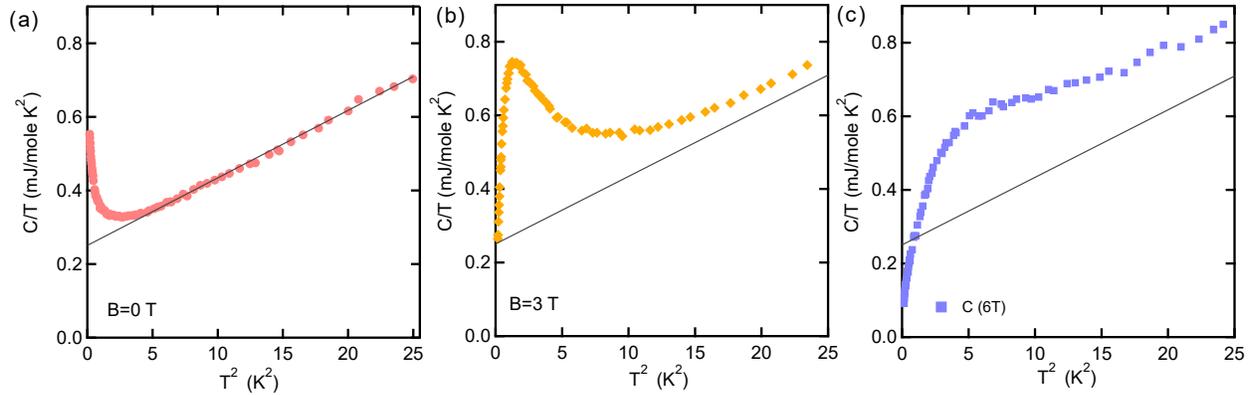

*Fig. 2. Specific heat of the $Ti_{9.5}Si_{90.5}$ amorphous alloy presented in the form $C/T$ versus $T^2$ at indicated magnetic fields 0, 3 and 6 Tesla. In all panels the solid line indicates a fit to zero-field specific heat in the range 2.5-5 K.*

At temperatures below 2 K, zero-field specific heat displays an additional upturn contribution (Fig. 2a), which evolves into the Schottky anomaly when magnetic field is applied (Fig. 2b,c). The appearance and evolution of this excess contribution $\Delta C$ is very similar to the behavior observed in doped Si:(P,B) semiconductors [9]. In both systems, the excess specific heat originates from magnetic moments; however, there is an important difference between the two cases.

In doped crystalline semiconductors, the magnetic moments are due to dopants themselves; for example, an isolated unionized phosphorus dopant carries spin ½. At small dopant concentration, spins interact with their neighboring spins via the exchange interaction. Because of the random locations of the dopants in the Si lattice, the exchange constants of this interaction have a very broad distribution. The Bhatt-Lee model gives theoretical description of magnetic and thermodynamic response of such random spin systems and reproduces the functional behavior $\Delta C \sim T^{-\alpha}$ observed experimentally in zero magnetic field. At high dopant concentrations close to the MIT, most of the electrons form a band and are non-magnetic; still, a small fraction of dopants well-isolated from the rest of the system continue to carry isolated magnetic moments.

The magnetic properties of Ti-Si (and V-Si) amorphous alloys were carefully studied in Ref. [29]. The combined magnetization and electron-spin resonance (ESR) measurements convincingly showed that the magnetic moments are associated not with isolated Ti and V atoms but with dangling bonds of amorphous silicon. The concentration of the spins was quantitatively determined and was found to grow with decreasing Ti content, concurrent with the growth of the portion of the system that has the local structure of amorphous silicon. The upturn in specific heat seen in zero magnetic field likely corresponds to the interaction between some of these spins. The ESR parameters of the spins are the same as in amorphous Si and not affected by Ti atoms, therefore, at least in the first approximation, we can treat them as an independent subsystem that stays away from conducting part of the system and does not affect the electronic properties of the alloys. This assertion is supported by the observation that in amorphous Si made by sputtering in argon atmosphere, as it is in our case, the dangling bonds appear on the walls of small cavities formed during the growth process [32].

The concentration of the magnetic moments in $Ti_{9.5}Si_{90.5}$ was found to be $n_s \approx 0.7 \times 10^{19}$ cm$^{-3}$. We also found that a field of 6 T overcomes any interaction between the spins, so at this field, the magnetic contribution should be described by the simple formula for the Schottky anomaly, $\Delta C = n_s k(\varepsilon/kT)^2 \exp(\varepsilon/kT)/[\exp(\varepsilon/kT)+1]^2$, with $\varepsilon = 2\mu_B B$. We computed this contribution for $B = 6$ T and subtracted it from the total specific heat shown in Fig. 2c.

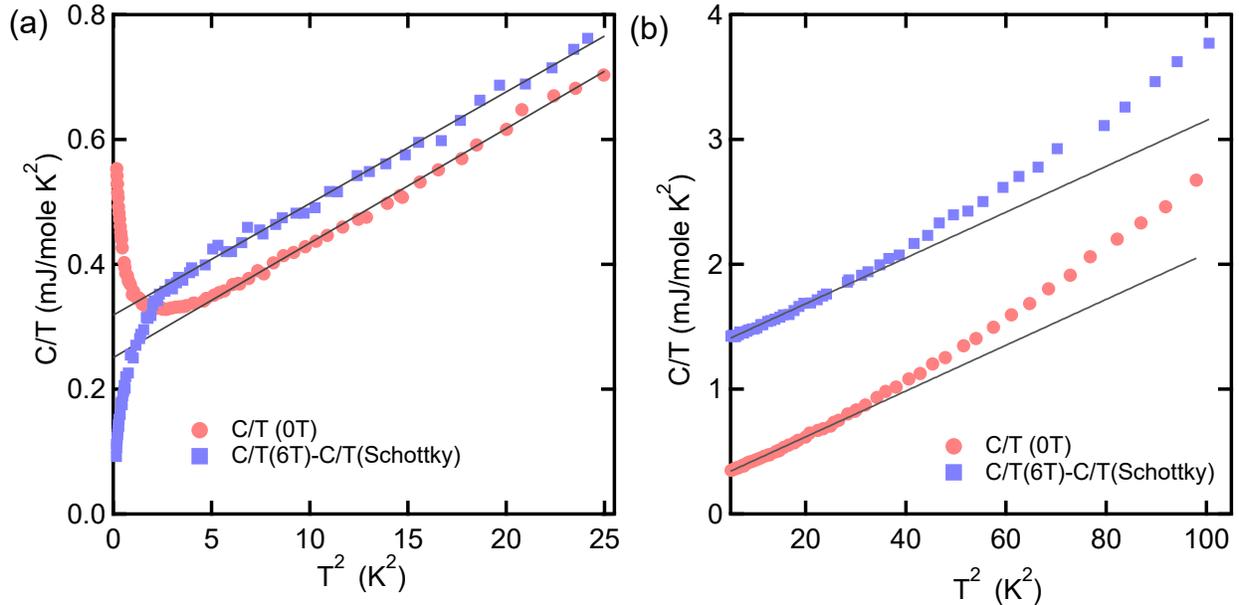

*Fig. 3 (a) Specific heat in zero magnetic field and specific heat at the field of 6 Tesla after subtracting contribution of magnetic moments. The solid lines are the linear fits to the data at temperatures 2.2-5 K. (b) Same quantities at higher temperatures. Data at 6 Tesla are shifted up by 1 mJ/mole K².*

The resulting specific heat at 6 Tesla, which combines electronic and photonic terms, is shown in Fig.3a,b alongside the data for zero field. Both data sets vary linearly with the same slope in the temperature range 2-5 K and show the same deviation from $T^2$ variation at higher temperatures. This deviation corresponds to the non-Debye lattice specific heat frequently observed in semiconductors [33]. We see, as expected, that phonon contribution is not affected by magnetic field. It is an important observation since it validates our subtraction procedure. The effect of magnetic field on the specific heat above 2 K thus corresponds to a simple increase of the electronic coefficient γ by about 20%. Below 2 K, specific heat starts to display an anomalous behavior and drops with decreasing temperature. This feature is not an artefact of our subtraction procedure; it is clearly present already in the raw data shown in Fig. 2c.

To characterize the anomalous behavior, we have subtracted the phonon contribution from the specific heat. The obtained temperature variation of the electronic specific heat coefficient $\gamma$ is shown in Fig.4 a,b; it constitutes the central result of our study. From this observation, we can make a claim that $\gamma$ displays critical behavior at the MIT of Ti-Si alloys. We tried to fit $\gamma(T)$ with several common dependences. Fig. 4a shows the fit to the power-law function and Fig.4b to logarithmic dependence $\gamma(T) = \gamma_0 \ln(T/T_0)$. The latter provides the best phenomenological approximation to our data. It is marked by two temperatures, $T_1 \approx 1.5$ K, below which $\gamma$ starts to drop and $T_0 \approx 0.2$ K, at which it apparently reaches zero.

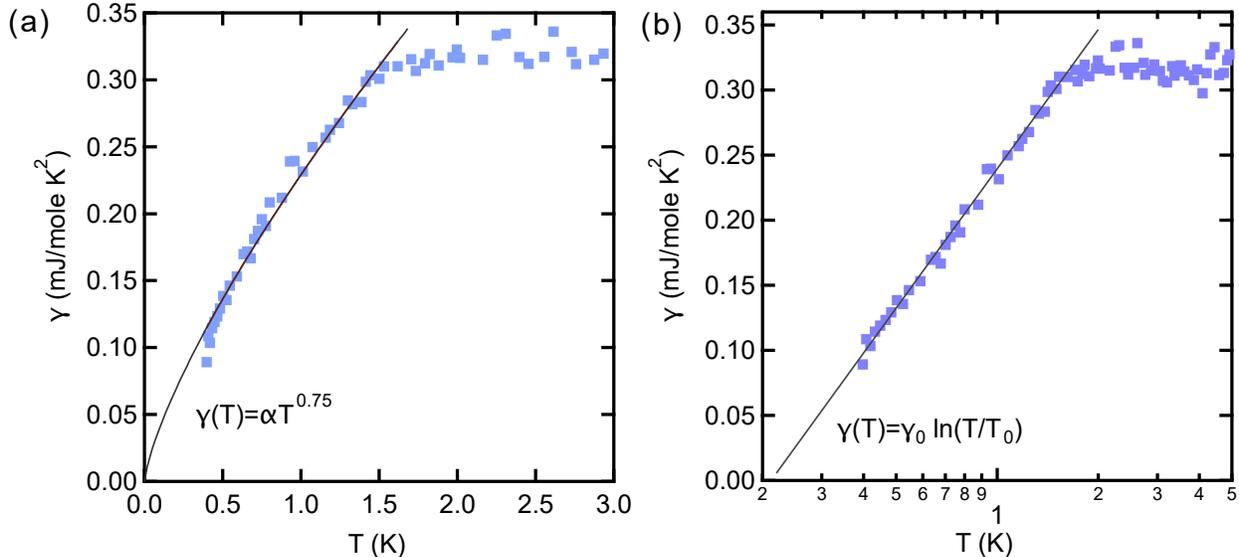

*Fig. 4. The electronic specific heat coefficient γ at magnetic field 6 Tesla as a function of temperature on linear (a) and logarithmic (b) scale. The solid lines show fits with the dependences indicated in the figure.*

The $Ti_{9.5}Si_{90.5}$ alloy displays a very unusual variation of electronic specific heat. As a starting point of the discussion of its possible origin, let us describe several relevant properties of the system. (1) It is very unlikely that the anomalous downturn is due to the superconducting gap. There is no evidence that Ti-Si amorphous alloys are superconducting. Titanium disilicide, which might hypothetically precipitate in the alloy, does not superconduct down to at least 50 mK [34]. (2) In our study of the Hall coefficient presented in a companion paper [35], we show that the EEI correction to this quantity is very large and observed up to the temperature 150 K (in doped semiconductors corresponding temperature is about 1 K). Within the perturbation theory, the EEI correction to Hall coefficient comes from the same processes that produce the anomaly in the tunneling single-particle density of states (spDOS). Therefore, we can consider $T_{sp} \approx 150$ K to be the energy scale of the EEI as seen by the spDOS. (3) In $Ti_{9.5}Si_{90.5}$, $\gamma$ measured above 2 K gives the density of states $g(E_F) \approx 5 \times 10^{21}$ eV$^{-1}$cm$^{-3}$ and, assuming for the sake of an estimate that effective mass is equal to the free electron mass, carrier concentration $n \approx 2 \times 10^{21}$ cm$^{-3}$. This is close to the concentration of Ti atoms, $n_{Ti} = 4.7 \times 10^{21}$ cm$^{-3}$, so EEI does not seem to affect $\gamma$. (4) Consistent with the Hall studies, the insulating Ti-Si alloys [28] show Efros-Schlovskii VRH at high temperatures, which surprisingly switches to Mott VRH below about 30 K. A close inspection of the data for V-Si (Fig. 2a in Ref. 11) indicates similar behavior. In insulating Ti-Si, the density of states extracted from magnetoresistance at $T$=4, 8 K within the model of non-interacting electrons was found to be two orders of magnitude smaller than the density of states extracted from specific heat [28].

Let us further notice that, in Mo-Ge [10], V-Si [11] and Ti-Si [12] amorphous alloys, the free-electron-like specific heat of roughly the same magnitude appears at 1.5 K $< T \ll T_{sp}$ and varies smoothly across the transition. This by itself is a puzzling observation, the origin of which is still not understood. Some attempts have been made to link it to the local atomic structure of the alloys. The structure of insulating V-Si alloys was determined by the neutron diffraction and corresponds, as sketched in Fig. 5a, to V atoms randomly distributed in amorphous Si matrix. We expect a similar arrangement in Ti-Si. The atomic structure of the insulating $Mo_xGe_{100-x}$ alloys ($x \leq 11$) was determined by anomalous small-angle and differential x-ray scattering techniques [36,37]. It is sketched in Fig. 5b and can be seen as a collection of metallic amorphous particles with the composition of intermetallic compound $MoGe_2$ and size $a_p \approx 1.5$ nm embedded in the matrix of amorphous Ge. The author of the study proposed that these metallic clusters are responsible for finite $\gamma$ in the alloys. We disagree with this conclusion. Using the data in Ref. [10] we estimate the density of state in amorphous $MoGe_2$ as $g_{MG2} \approx 5 \times 10^{22}$ eV$^{-1}$cm$^{-3}$ and the level spacing in the particles as $\Delta E \approx 1/g_{MG2} a_p^3 \approx 60$ K. A collection of *isolated* particles of this kind would produce not the free-electron-like specific heat but rather a response akin to the Schottky anomaly with a peak at 60 K. To have finite $\gamma$, the electron wave functions need to extend over many particles, and in this regard, their behavior is not conceptually different from wave functions in V-Si.

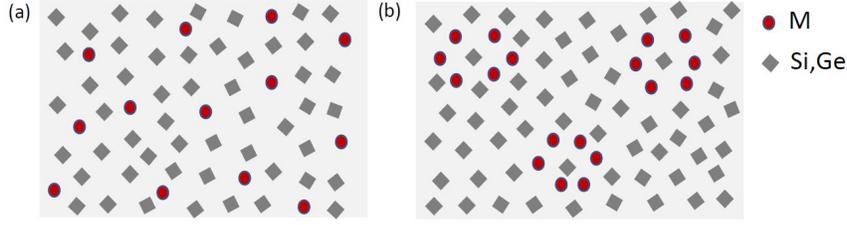

*Fig. 5. Schematic representation of local atomic structure of amorphous metal – Si,Ge alloys, at low metal concentration. (a) Metal atoms are randomly distributed in the amorphous Si,Ge matrix. (b) Metal atoms form amorphous clusters with a local structure of an intermetallic crystalline compound $M(Ge,Si)_2$ in the matrix of amorphous (Si,Ge).*

To summarize, the specific heat in Ti-Si, V-Si and Mo-Ge amorphous alloys cannot be explained by the clustering and emerges deep inside the temperature range where EEI affects the spDOS. In our view, it should be related to the many-electron excitations. Let us now compare our results with existing theories.

The behavior of specific heat in several universality classes was studied by Castellani and Di Castro within the generalized nonlinear σ model [38]. Two of their predictions look similar to our observations. In the case of strong magnetic fields ($\mu_B B \gg kT$), the predicted $\gamma$ is rescaled but does not acquire any temperature dependence. In the spin-orbit impurity case, $\gamma$ approaches zero at the MIT as $\gamma \sim T^{1/2}$. However, the interpretation based on this theory also raises concerns and should explain why experimental and theoretical temperature exponents are different, why spin-orbit behavior is dominant at low temperatures, and why the rescaling predicted for $\mu_B B \gg kT$ actually persists to much higher temperatures.

The second model that captures some of our observations is the model of the Coulomb glass. Several theoretical studies [39,40,41,42] of this state predict finite $\gamma$ caused by many-electron excitations coexisting with a large Coulomb gap in the spDOS. Moreover, temperature-dependent $\gamma$ has been predicted to occur at lowest temperatures, similar to the anomalous behavior in Ti-Si below 2 K. However, we have not observed the decrease of $\gamma$ at high temperatures universally predicted by these theories. Also, in our specific heat and low-bias resistance measurements, we did not observe any relaxation or aging effects of the type reported for indium oxide [43] and considered to be inherent for the Coulomb glass [44].

The third possibility is that our observations reflect the response of the many-body localized state. The evidence is all indirect: the $Ti_{9.5}Si_{90.5}$ alloy behaves as if a larger and larger portion of electrons gets decoupled from the environment with decreasing temperature; the experimental dependence of $\gamma$, $\gamma(T) = \gamma_0 \ln(T/T_0)$ (if it continues) gives zero $\gamma$ at finite temperature $T_0 \approx 200$ mK; and the dependence of $\gamma$ curiously matches logarithmic dynamics of entropy found in some MBL studies [14,25]. Interestingly, when the author of Ref. [24] discussed possible experimental manifestation of MBL, they brought as an example bistable $I-V$ curves observed in an amorphous $Y_{20}Si_{80}$ alloy below 50 mK [45]. This system belongs to the same class of materials as Ti-Si alloys and similar to our $Ti_{9.5}Si_{90.5}$ sample, $Y_{20}Si_{80}$ is located in the immediate vicinity of MIT.

*Outlook*

We have found that the electronic specific heat in a Ti-Si amorphous alloy located very close to the critical point of the MIT displays an anomalous logarithmic downturn at low temperatures. The observed behavior of γ can possibly be related to the criticality at the MIT or to the formation of the Coulomb glass or many-body localized state. While we believe that the anomalous behavior is likely caused by MBL, more evidence is needed to fully accept or dismiss any of these three pictures.

The presence of the Coulomb glass and MBL states in condensed matter systems is not currently established. In our view, Metal-(Si,Ge) amorphous alloys present a versatile class of materials very suitable for exploration of both of these phenomena and perhaps for disentangling the differences between them. Let us mention a few advantages of these systems. The alloys can be easily fabricated by sputtering at room temperature. The crystallization temperature is fairly high (600°C for $Ti_{9.5}Si_{90.5}$ alloy; see Ref. 11 for more details) and, from our experience, the properties of the samples do not change for 2-3 years. In a longer time, the alloys are likely to experience the phase separation shown in Fig. 5 and consequent growth of the metallic clusters. Same clustering processes can be induced by heat treatment; indeed, in a few test experiments, we found that annealing of alloys below their crystallization temperature leads to a significant counterintuitive growth in resistance. The phase-separated structure can be fairly easily characterized by dispersive X-ray and neutron techniques. Accidentally, this structure closely resembles the disorder pattern theoretically suggested to test thermalization effects in MBL (see Fig. 12 in Ref. 14). The thin films of the amorphous alloys provide model 2d systems. They can also be patterned using negative resist e-beam lithography to create sub-10 nm 1d nanowires and a variety of other custom-designed structures [46]. This is a very important capability since most of the work on MBL addresses 1d systems. Let us finally mention that the state-of-the-art heat capacity techniques, such as those used in the studies of 2d-gas in GaAs-based quantum wells [47], allow one to reach temperatures ($\sim 25\ mK$) below the apparent glass or MBL transitions. Moreover, application of a sufficiently large magnetic field (18 T) would shift the Schottky anomaly to high temperatures and leave the electronic contribution unobscured below $T \approx 5$ K. To conclude, we hope that further, more focused exploration of the amorphous alloys will clarify the physics of the Coulomb glass and MBL.

## Acknowledgements

AR gratefully acknowledges the support by NSF Grant DMR1904221.